\documentclass[leterpaper]{aastex}
\usepackage{multirow}
\usepackage{float}
\usepackage{natbib}

\slugcomment{To be submitted to the Astrophysical Journal (ApJ)}
\shorttitle{Hot NH$_{3}$ Spectra}
\shortauthors{Hargreaves et al.}

\begin{document}

\title{HOT NH$_{3}$ SPECTRA FOR ASTROPHYSICAL APPLICATIONS}

\author{\textsc{Robert J. Hargreaves}
\affil{Department of Chemistry, University of York, Heslington, York, YO10 5DD,
UK; rjh135@york.ac.uk}}

\author{\textsc{Gang Li}
\affil{Department of Chemistry, University of York, Heslington, York, YO10 5DD,
UK; gl525@york.ac.uk}}

\and

\author{\textsc{Peter F. Bernath}}
\affil{Department of Chemistry, University of York, Heslington, York, YO10 5DD,
UK; pfb500@york.ac.uk}

\begin{abstract}

We present line lists for ammonia (NH$_{3}$) at high temperatures obtained by recording Fourier transform infrared emission spectra. Calibrated line lists are presented for twelve temperatures (300 -- 1300$^{\circ}$C in 100$^{\circ}$C intervals and 1370$^{\circ}$C) and each line list covers the 740 -- 2100 cm$^{-1}$ range, which includes the majority of the $\nu_{2}$ umbrella bending mode region (11 $\mu$m) and the $\nu_{4}$ asymmetric bending mode region (6.2 $\mu$m). We also demonstrate the useful technique of obtaining empirical lower state energies from a set of spectra recorded at different sample temperatures. Using our NH$_{3}$ spectra, we have estimated lower state energies ($E_{Low}$ in cm$^{-1}$) and our values have been incorporated into the line lists along with calibrated wavenumbers ($\tilde{\nu}$ in cm$^{-1}$) and calibrated line intensities ($S'$ in cm/molecule) for each line. We expect our hot NH$_{3}$ line lists to find direct application in the modelling of planetary atmospheres and brown dwarfs.

\end{abstract}


\section{INTRODUCTION}

High-resolution infrared laboratory spectra are an essential element in numerous fields of science such as atmospheric chemistry \citep{ACE05} and astronomy \citep{swain08}. These high-resolution spectra allow molecular emission and absorption features to be accurately identified and assigned; for example the detection of `Water on the Sun' \citep{wallace95} was made possible by the use of high-resolution hot water (H$_{2}$O) line lists derived from experimental spectra. The H$_{2}$O molecule is able to exist in sunspots (typically 3,000 -- 4,500 K), and observations showed that many previously unassigned lines were in fact due to the ubiquitous H$_{2}$O molecule.

Molecules form in `cool' sources \citep{bernath09} and progress in infrared astronomy has led to the identification of numerous molecules in cool stars \citep{wingford69,allard97}, brown dwarfs \citep{geballe02,sharpeburrows07} and, most recently, in exoplanets \citep{swain09,seagerdeming10}.

After H and He, C, N and O are the most abundant elements, so in cool objects one anticipates that methane (CH$_{4}$), NH$_{3}$, and H$_{2}$O will be abundant. Although the spectra of the fundamental infrared bands of `cold' (i.e. room temperature) CH$_{4}$, NH$_{3}$, and H$_{2}$O have been understood for many years, the situation for overtones, hot bands and combination bands is not so sanguine. For H$_{2}$O it is only recently that high resolution experimental and theoretical line lists suitable for computing the spectral energy distributions (SEDs) of astronomical objects with temperatures in the 300 -- 4000 K range have become available \citep{coheur05,barber06}. For `hot' CH$_{4}$, there are only some rather sparse experimental observations and no satisfactory quantitative theoretical understanding. For hot NH$_{3}$, there has been some recent theoretical progress \citep{yurchenko11, zobov11, huang11a, huang11b} and we report here on an extensive set of experimental observations in the thermal infrared.

The NH$_{3}$ molecule is one of the simplest polyatomic molecules with four atoms forming a trigonal pyramidal structure with C$_{3v}$ symmetry \citep{bernath05}. NH$_{3}$ has $3N-6=6$ fundamental vibrational modes, of which two are doubly degenerate. There are therefore four fundamental vibrational frequencies $\nu_{1}(a_{1})$ at 3336.2 cm$^{-1}$ (N-H stretch), $\nu_{2}(a_{1})$ at 932.5 cm$^{-1}$ (umbrella mode), $\nu_{3}(e)$ at 3443.6 cm$^{-1}$ (N-H stretch) and $\nu_{4}(e)$ at 1626.1 cm$^{-1}$ (bend), of which $\nu_{3}$ and $\nu_{4}$ are degenerate (Table \ref{tab1}). There is an additional complication in the spectroscopy of NH$_{3}$ due to the low barrier to `inversion'. Although NH$_{3}$ is non-planar at equilibrium the three hydrogen atoms can flip through the plane containing the N atom and perpendicular to the C$_{3}$ symmetry axis.

\begin{deluxetable}{lcrrl|lrl}
\tabletypesize{\small}
\tablenum{1}
\label{tab1}
\tablewidth{0pt}
\tablecaption{The vibrational modes and rotational constants of NH$_{3}$ \citep{herzberg91}.}
\tablehead{\colhead{Mode} & \colhead{Symmetry} & \multicolumn{2}{c}{Frequency (cm$^{-1}$)} & \colhead{Type} & \multicolumn{2}{c}{Rotational Constants}}
\startdata
$\nu_{1}$ & $a_{1}$ & 3336.2 & (3337.2\tablenotemark{*})  & Symmetric Stretch        & $B_{0}$   & 9.9443 cm$^{-1}$\\
$\nu_{2}$ & $a_{1}$ &  932.5 &  (968.3\tablenotemark{*})  & Symmetric Bend (Umbrella)& $C_{0}$   &  6.196 cm$^{-1}$\\
$\nu_{3}$ &  $e$    & 3443.6 & (3443.9\tablenotemark{*})  & Asymmetric Stretch   & $r_{0}$   & 1.0173 {\AA} \\
$\nu_{4}$ &  $e$    & 1626.1 & (1627.4\tablenotemark{*})  & Asymmetric Bend      & $\alpha$  &  107.8 $^{\circ}$ \\
\enddata
\tablenotetext{*}{Molecular inversion causes the modes to be doubled.}
\end{deluxetable}

\clearpage

If the H atoms were labeled this would correspond to a right-handed form of the molecule converting into a left-handed form. Hence, there are two forms (frameworks) of the NH$_{3}$ molecule (energy levels are conventionally labeled as $s$ and $a$ or + and -) which can interconvert. Indeed, the interconversion frequency at 24 GHz ($s-a$ transition for $J=1$, $K=1$) can be used by radio astronomers to monitor NH$_{3}$ in dark interstellar clouds \citep{benmyers89}. In general, this inversion motion doubles vibration-rotation lines (each of the two interconverting forms has slightly different energy levels) and leads to the two slightly different band origins listed in Table \ref{tab1}.

NH$_{3}$ is an oblate symmetric top ($B=9.9443$ cm$^{-1}$, $C=6.196$ cm$^{-1}$) and the fundamental modes are either parallel ($\nu_{1}$ and $\nu_{2}$, $a_{1}-a_{1}$) or perpendicular ($\nu_{3}$ and $\nu_{4}$, $e-a_{1}$) depending on whether the infrared transition dipole moment is oscillating parallel or perpendicular to the C$_{3}$ symmetry axis. The rotational quantum numbers $J$ and $K$, describe the quantization of the total angular momentum (excluding nuclear spin) and its projection onto the symmetry axis, respectively. Rotational selection rules ($\Delta K=0$, $\Delta J=0,\pm1$) for a parallel transition lead to a simple $P$, $Q$, $R$ structure ($\Delta J=-1,0,1$) with a small $K$ splitting. The selection rules for perpendicular transitions ($\Delta K=0,\pm1$, $\Delta J=0,\pm1$) lead to a more complex pattern with widely spaced $K$ structure, organized into $K$ sub-bands each with different $K^{\prime}-K^{\prime\prime}$ values and $P$, $Q$, $R$ structure \citep{bernath05}.

 Over the years, numerous studies have been undertaken to assign the complicated NH$_{3}$ spectrum \citep{pandm83, pandm84, sasada92, kleiner99, cottaz00, lees08}. Many studies are of cold NH$_3$ and are useful when studying planetary atmospheres in our solar system \citep{brownmarg96}. Some work has been done in determining empirical lower state energies \citep{brownmarg96} as well as on the rotation-inversion spectra \citep{sasada92, yu2010}. Recently there has been considerable progress in the variational calculation of vibration-rotation energy levels and transitions from adjusted ab initio potential energy surfaces \citep{huang2008, huang11a, huang11b, yurchenko09, yurchenko11}. Calculated line lists suitable for spectra of ammonia at 300 K \citep{yurchenko09} and 1500 K \citep{yurchenko11} are now available.

Cool objects are less luminous than hot stars and emit most of their radiation in the infrared. High resolution infrared spectra of these objects are best recorded with large 10 m class telescopes. In 1995 the first brown dwarf, Gliese 229B was identified by the presence of methane in its spectrum \citep{oppen95}. Objects with masses less than 0.0075 solar masses are unable to fuse hydrogen in their cores \citep{burrows01} and are called brown dwarfs. Brown dwarf atmospheres are cool enough (typically 700 -- 2,000 K) to sustain a rich molecular environment \citep{kirkpatrick05}. Since the discovery of Gliese 229B, the L and T classes have been introduced to describe objects cooler than M-dwarfs \citep{leggett01}. L-dwarfs display near infrared electronic transitions of metal hydrides such as FeH \citep{hargreaves10} and CrH \citep{kirkpatrick99} and have weak TiO and VO transitions which are characteristic of M-dwarfs \citep{boch2007}. In contrast, T-dwarfs like Gliese 229B have prominent overtone transitions of hot H$_{2}$O and CH$_{4}$ \citep{burg2006}. Based on observations taken from the \textit{Spitzer Space Telescope}, \citet{cushing06} identified prominent H$_{2}$O, CH$_{4}$ and NH$_{3}$ absorption bands between 6 -- 14 $\mu$m, including the NH$_{3}$ $\nu_{2}$ band near 11 $\mu$m. It is predicted that a new cooler classification, a Y-dwarf ($<$ 700 K) is required below the T-dwarf class \citep{kirkpatrick05} with the prediction that strong NH$_{3}$ absorptions will help to differentiate the Y-dwarfs from T-dwarfs \citep{leg07, delorme08}.

\begin{figure}[h]
\figurenum{1}
\label{fig1}
\epsscale{1.0}
\plotone{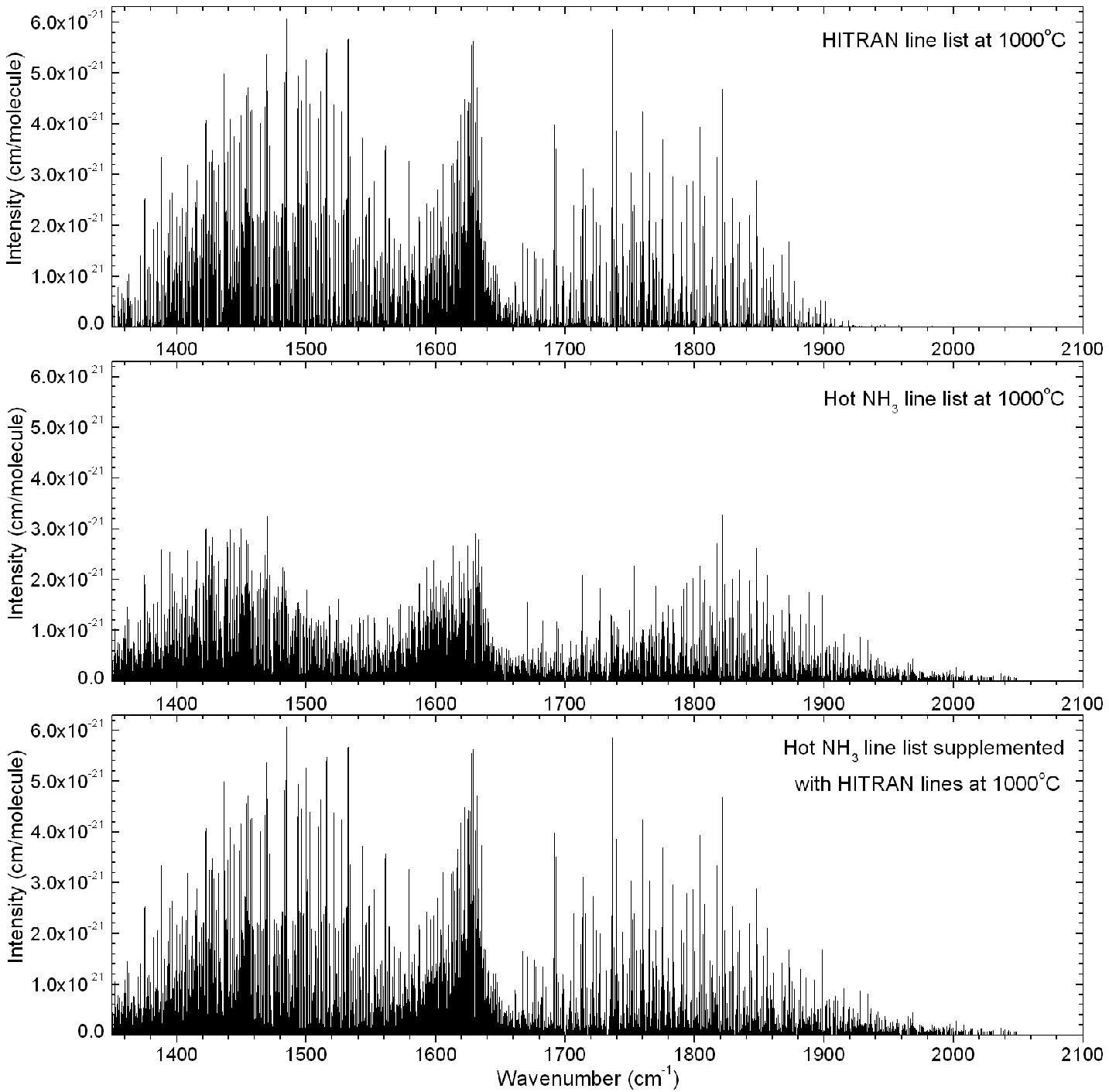}
\figcaption{A line spectrum comparison of NH$_{3}$ between the HITRAN line list converted to 1000$^{\circ}$C (top), the calibrated emission line list at 1000$^{\circ}$C (middle) and the final line list containing `hot' emission lines supplemented with HITRAN lines (with an intensity greater than $1\times10^{-22}$ cm/molecule) at 1000$^{\circ}$C (bottom).}
\end{figure}

Numerous unassigned features are present in the observed SEDs of brown dwarfs. Due to a lack of experimental data recorded in the temperature ranges appropriate for these objects (or reliable theoretical predictions), the best option is to extrapolate room temperature `cold' line parameters such as the HITRAN 2008 database \citep{rothman09} to the temperature of interest (typically 1000 K for sub-stellar brown dwarfs). For this reason, these extrapolated spectra rarely match observations satisfactorily (see Figure \ref{fig1}) due to the lack of hot bands and highly-excited rotational transitions which contribute significantly to the SED in these temperature ranges.

Along with brown dwarfs, it has only recently become possible to detect planets outside of the Solar System (exoplanets). 51 Pegasi b was the pioneering discovery by \citet{mandq95} that paved the way for many more exoplanet detections, with over 500 discovered to date (http://exoplanet.eu/). The vast majority of these exoplanets are large gas giants close to their parent star with high temperature gaseous atmospheres typically referred to as `hot Jupiters'. The new technique of `transit spectroscopy' \citep{char00} has been used to probe exoplanet atmospheres \citep{char02} and familiar molecules such as H$_{2}$O, CH$_{4}$, CO and CO$_{2}$ have all been shown to be present in examples like the well studied HD 209458b \citep{swain09,snellen10}.

The properties (composition, temperature/pressure profiles, etc.) of exoplanet atmospheres retrieved from the emergent flux or from transit spectroscopy depend on the molecular opacities used to simulate the observed spectra. Spectroscopic studies of brown dwarfs and exoplanet atmospheres therefore need improved line parameters for hot NH$_{3}$ and CH$_{4}$.

Currently the best line list for NH$_{3}$ is the HITRAN 2008 database, which is incomplete and has insufficient hot band information for the astronomical applications considered here. Moreover recent theoretical work has demonstrated that HITRAN 2008 has many errors in quantum number assignments of NH$_{3}$ \citep{huang11b, yurchenko09}. NH$_{3}$ has already been shown to contribute significantly to the SED of brown dwarfs at $\sim$620 K \citep{delorme08}. Improved theoretical predictions for the vibration-rotation lines will help to assign the complicated spectra of NH$_{3}$ \citep{zobov11} presented in this paper. We provide high temperature line lists for NH$_{3}$ using a similar approach to that of \citet{nassar03}; in addition we derive empirical lower state energies for many of the lines. The line lists can be used directly to model the SEDs of brown dwarfs leading to a better understanding of the T-/Y-dwarf boundary and perhaps allow the first identification of NH$_{3}$ in an exoplanet atmosphere.

\clearpage

\section{EXPERIMENTAL METHOD}

\begin{figure}[H]
\figurenum{2}
\label{fig2}
\epsscale{1.0}
\plotone{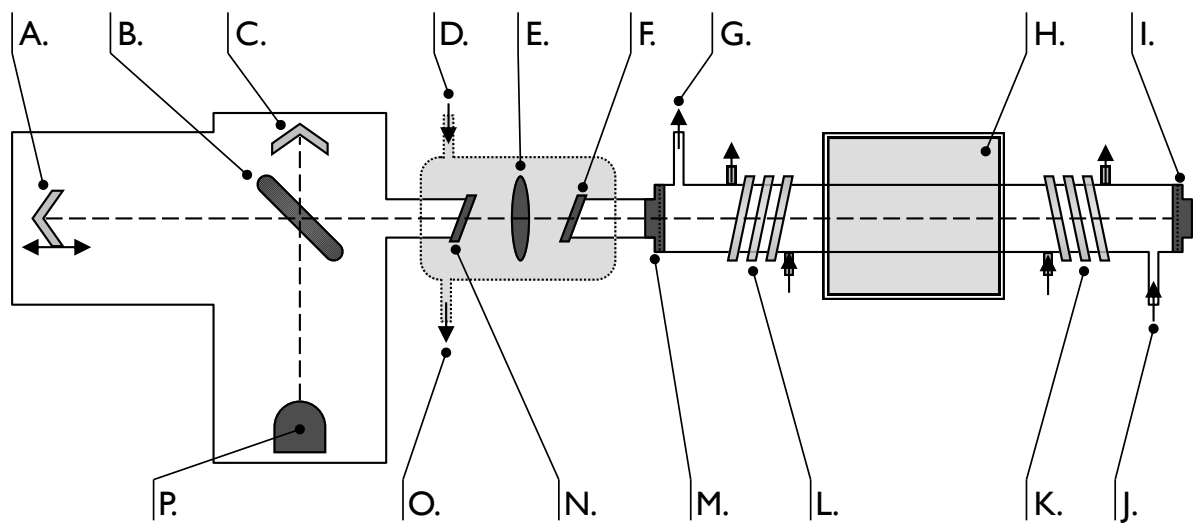}
\figcaption{A diagram of the experimental setup used (not to scale). The FT-IR spectrometer is on the left and the Al$_{2}$O$_{3}$ tube is on the right, the dashed line indicates the line of sight of the FT-IR spectrometer and the labels refer to:  A. Scanning corner-cube; B. Beamsplitter;  C. Fixed corner-cube;  D. Dry air purge;  E. Lens;  F. Window;  G. NH$_{3}$ exits to vacuum pump;  H. Tube furnace;  I. Stainless steel end cap with rubber o-ring;  J. NH$_{3}$ enters from cylinder;  K. and L. Cooling H$_{2}$O coils;  M. Stainless steel end cap with rubber o-ring;  N. Window;  O. Dry air exit; and  P. MCT detector.}
\end{figure}

Laboratory emission spectra of hot NH$_{3}$ were recorded using a system similar to that used by \citet{nassar03} to study hot CH$_{4}$. A diagram of the experiment is shown in Figure \ref{fig2}. An alumina (Al$_{2}$O$_{3}$) sample tube is sealed with windows and evacuated. A constant slow flow of NH$_{3}$ gas is introduced to the system and a constant pressure is maintained using a needle valve which helps to minimize the build up of impurities and loss of sample within the system. Surrounding the central 51 cm of the Al$_{2}$O$_{3}$ tube (121 cm long) is a controllable tube furnace capable of maintaining stable temperatures up to 1370$^{\circ}$C. The radiation exiting the Al$_{2}$O$_{3}$ tube is focussed into a Fourier transform infrared (FT-IR) spectrometer using a lens and the distance between the window on the end of the Al$_{2}$O$_{3}$ tube and the FT-IR spectrometer entrance aperture was 20 cm. This volume was purged with `dry air' (H$_{2}$O absent) to minimize the presence of H$_{2}$O absorption lines in the emission spectra.

\begin{figure}[H]
\figurenum{3}
\label{fig3}
\epsscale{1.0}
\plotone{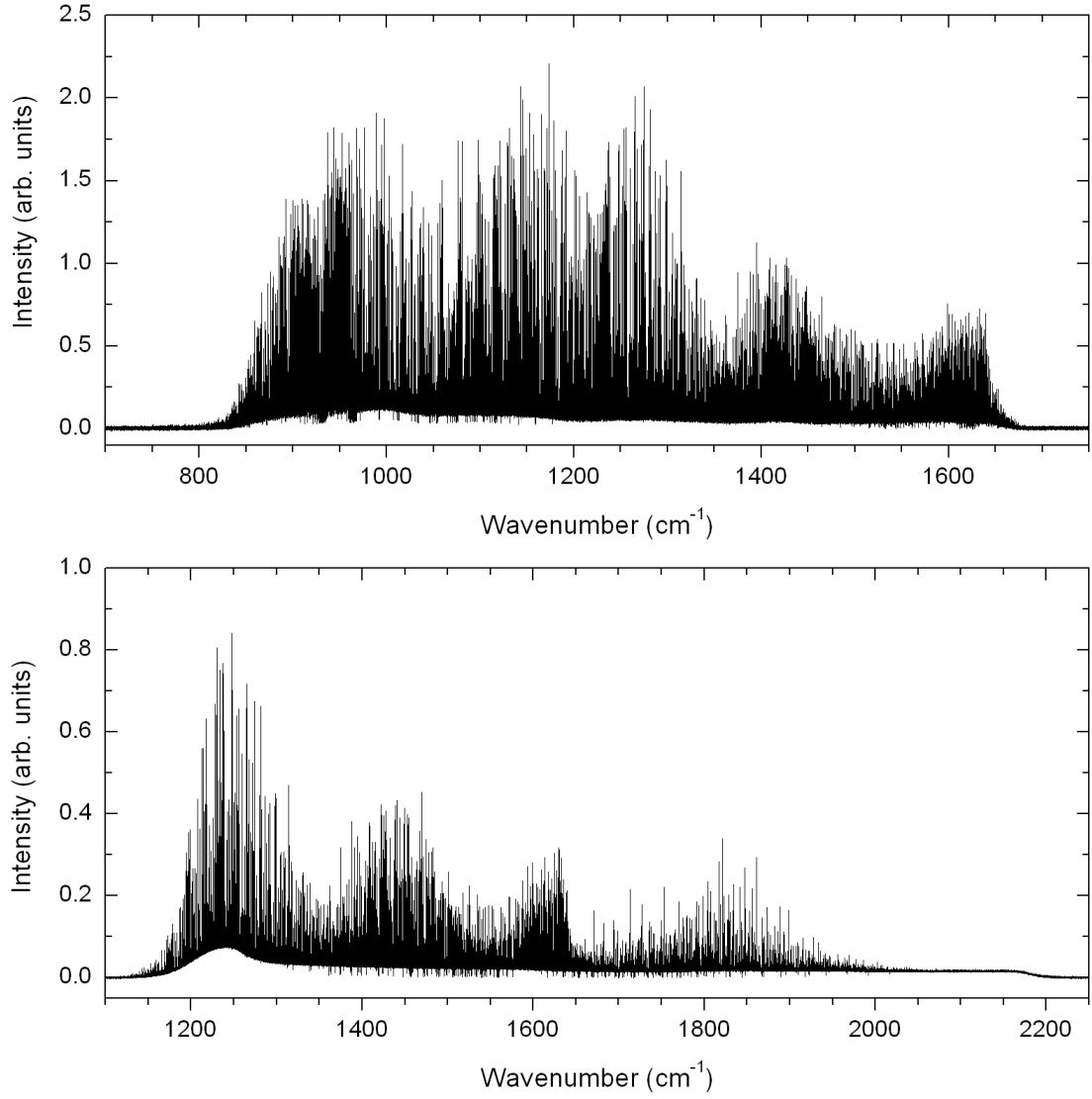}
\figcaption{The two observed infrared emission spectral regions of NH$_{3}$ at 1000$^{\circ}$C.}
\end{figure}

We recorded all spectra in two sections to improve the signal-to-noise ratio (see Figure \ref{fig3}) and the experimental conditions for each region are listed in Table \ref{tab2}. The two selected regions were a consequence of the available filters and constituent materials of the system. The first region covered the range 740 -- 1690 cm$^{-1}$ and was recorded with thallium bromoiodide (KRS-5) windows, a potassium bromide (KBr) beamsplitter, a zinc selenide (ZnSe) lens and a liquid nitrogen cooled mercury-cadmium telluride (MCT) detector. The lower wavenumber cut off was limited due to the band gap of the MCT detector and a small part of the NH$_{3}$ $\nu_{2}$ umbrella mode could not be observed.

\begin{deluxetable}{lrr}
\tabletypesize{\small}
\tablenum{2}
\label{tab2}
\tablewidth{0pt}
\tablecaption{Experimental conditions.}
\tablehead{\colhead{Parameter} & \colhead{Region 1} & \colhead{Region 2}}
\startdata
Spectral Region (cm$^{-1}$)  &   740 - 1690 &  1080 - 2200 \\
Detector                     &          MCT &          MCT \\
Beamsplitter                 &          KBr &          KBr \\
Windows                      &        KRS-5 &    CaF$_{2}$ \\
Lens                         &         ZnSe &    CaF$_{2}$ \\
Scans                        &          240 &           80 \\
Resolution (cm$^{-1}$)       &         0.01 &         0.01 \\
Aperture (mm)                &         3.15 &          2.5 \\
NH$_{3}$ pressure (Torr)     &          5.0 &          1.0 \\
Zerofilling factor           &   $\times16$ &   $\times16$ \\
\enddata
\end{deluxetable}

\clearpage

The second region covered the range 1080 -- 2200 cm$^{-1}$ (although no lines were observed above 2100 cm$^{-1}$) and was recorded with calcium fluoride (CaF$_{2}$) windows, a KBr beamsplitter, a CaF$_{2}$ lens and liquid nitrogen cooled MCT detector. This second region covered the NH$_{3}$ $\nu_{4}$ bending mode, plus the R-branch of the $\nu_{2}$ mode. For both regions, 11 spectra were recorded at 100$^{\circ}$C intervals between 300$^{\circ}$C and 1300$^{\circ}$C. The twelfth spectrum was recorded at 1370$^{\circ}$C because the furnace had difficultly reaching 1400$^{\circ}$C. The resolution was chosen based on the pressure and Doppler broadening widths and the final line list was created by cutting both regions at 1350 cm$^{-1}$ and combining them to make a list of lines ranging from 740-2100 cm$^{-1}$ for each temperature.

A thermocouple within the interior of the tube furnace combined with a programmable controller was used to measure the furnace temperature and maintain high temperatures over long time periods. It was necessary to water cool the ends of the Al$_{2}$O$_{3}$ tube to avoid damaging the rubber o-rings at both ends of the tube which maintained the pressure seal. As a result, the ends of the Al$_{2}$O$_{3}$ tube are at a lower temperature than the central portion and because of the strong temperature dependence, the observed infrared emission spectra are dominated by the high temperature central portion of the heated sample.

All spectra contain emission lines of NH$_{3}$ assumed to be at the furnace temperature. The baseline of each spectrum depends upon the temperature and arrangement of the system; it originates primarily from thermal emission from the Al$_{2}$O$_{3}$ tube walls. The spectra also contain a small number of absorption lines due to H$_{2}$O vapor (see Figure \ref{fig3}). We also observed absorption of strong lines due to cold NH$_{3}$ in the beam path, mainly in the end of the tube nearest the spectrometer. A description of how we deal with these absorption lines is given below.

Once the spectra had been recorded, the peak-picking programme WSpectra \citep{carleer01} was used to identify all of the emission peaks at each temperature; the number of lines in each file is given in Table \ref{tab3}. The same program was then used to fit a Voigt profile to every emission line to create a line list (wavenumber, arbitrary intensity) for each temperature. The lines in the overlapping 1080 -- 1690 cm$^{-1}$ portion were compared at each temperature and calibrated to Region 1. It was decided that the two regions would be joined at 1350 cm$^{-1}$ as there is a natural minimum in the line emissions and this value was approximately in the center of the overlapping section. Hence from here onwards, Region 1 contains lines in the range 740 -- 1350 cm$^{-1}$ and Region 2 contains the lines 1350 -- 2100 cm$^{-1}$.

\begin{deluxetable}{lcc}
\tabletypesize{\small}
\tablenum{3}
\label{tab3}
\tablewidth{0pt}
\tablecaption{The number of measured emission lines.}
\tablehead{Temperature ($^{\circ}$C) & \colhead{Region 1\tablenotemark{a}} & \colhead{Region 2\tablenotemark{a}}}
\startdata
300    &    2,924 &   4,014  \\
400    &    4,280 &   4,885  \\
500    &    5,796 &   7,186  \\
600    &    7,665 &   7,433  \\
700    &    8,922 &   9,810  \\
800    &    8,841 &   8,562  \\
900    &   10,090 &  10,851  \\
1000   &   12,712 &   9,526  \\
1100   &   12,665 &   9,752  \\
1200   &   11,174 &   9,992  \\
1300   &    8,929 &   9,233  \\
1370   &    7,405 &   6,366  \\
\enddata
\tablenotetext{a}{The upper (lower) limit of region 1 (2) was chopped at 1350 cm$^{-1}$.}
\end{deluxetable}

\clearpage

\begin{figure}[H]
\figurenum{4}
\label{fig4}
\epsscale{1.0}
\plotone{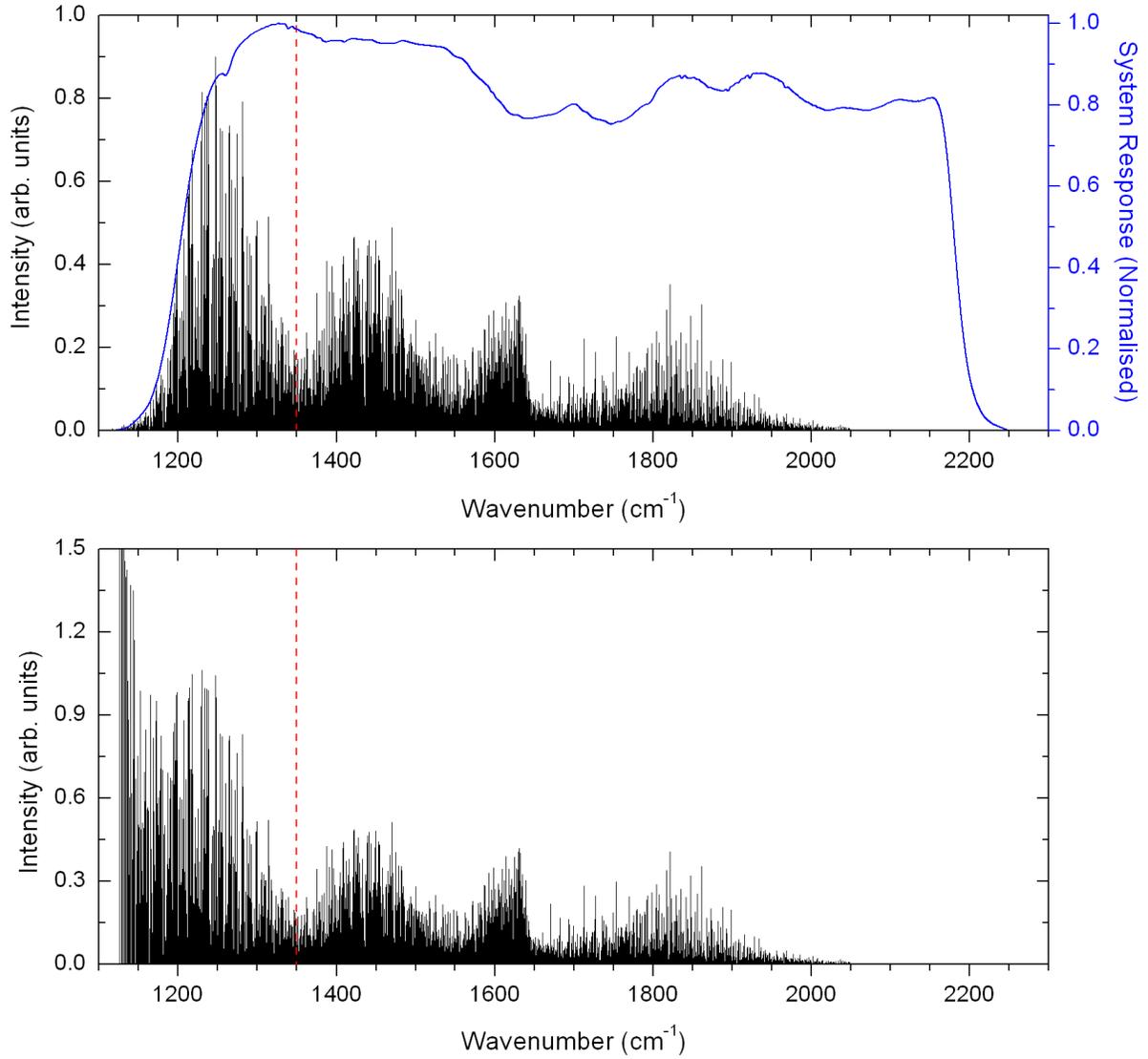}
\figcaption{The top panel shows an observed uncalibrated NH$_{3}$ line spectrum at 1000 $^{\circ}$C and the system response (blue) for region 2. The red dashed line indicates the low wavenumber cut-off for this region (i.e. 1350 cm$^{-1}$). The bottom panel shows the resulting lines corrected for system response.}
\end{figure}

A system response function for both regions was needed to account for the contribution from the system setup (e.g. windows, beamsplitter, lens, filter and detector) which alters the strength of the lines observed. This was achieved by recording the blackbody spectrum emitted from a solid graphite rod placed at the center of the Al$_{2}$O$_{3}$ tube; the end facing the spectrometer was machined into a concave cone to allow only blackbody emissions to be detected. The spectrum was then compared with a theoretical blackbody spectrum at the same temperature and normalized. The detector response curve for Region 2 and the effect on the line intensities can be seen in Figure \ref{fig4}.

All 24 fitted line lists (12 for each region) were wavenumber calibrated by comparison to 20 clean lines in the 2008 HITRAN database \citep{rothman09}. Strong, clear symmetric lines were favored and a calibration factor was then applied to correct the wavenumber scale. A typical calibration factor (obtained by dividing the HITRAN line center by observation) for region 1 was 1.000001739 (from the 900 $^{\circ}$C line list)  resulting in a shift of 0.00209 cm$^{-1}$ at 1200 cm$^{-1}$ and a typical calibration factor for region 2 was 1.000000990 (from the 900 $^{\circ}$C line list) resulting in a shift of 0.00158 cm$^{-1}$ at 1600 cm$^{-1}$. The overall accuracy of our wavenumber scale is better than $\pm$0.002 cm$^{-1}$ after calibration.

\section{THEORY}

Our analysis method is not unique and has successfully been applied by many workers, most recently in the sub-millimeter regime to study astrophysical `weeds' \citep{fortman10}. It involves a comparison of the observed intensities to a reliable data set (in our case, the 2008 HITRAN database). In order to make our emission line lists comparable to HITRAN we converted them into absorption intensities \citep{nassar03}. The relationship between emission and absorption intensities is given by
\begin{equation}
S_{absorption} = \frac{ S_{emission} }{ \nu^{3} \exp\left( -\frac {h \nu} {k T} \right)},
\end{equation}
where $S_{emission}$ is the intensity of the emitted line, $\nu$ is the frequency and $T$ is the temperature.

The HITRAN database provides line intensities at 296 K so they were converted to the relevant temperatures of the fitted line lists. The intensity of a line ($S'$) is defined \citep{bernath05} in SI units as
\begin{equation}
S' = \frac{ 2 \pi^{2} \nu_{10} S_{J'J''}}{3 \varepsilon_{0} h c Q} \exp\left(-\frac {E_{Low}} {k T}\right) \left[1 - \exp\left(-\frac {h \nu_{10}} {k T}\right)\right],
\end{equation}
where $S_{J'J''}$ is the square of the transition dipole moment, $Q$ is the total internal partition function and $E_{Low}$ is the lower state energy of the line. If the only changing variable is the temperature, then the equation can be used to obtain the intensity of the same line at any new temperature by dividing by a reference intensity ($S'_{0}$), giving
\begin{equation}
\frac{S'}{S'_{0}}= \frac{Q_{0}}{Q} \exp\left(\frac {E_{Low}} {k T_{0}} - \frac {E_{Low}} {k T}\right) \left[\frac{1 - \exp\left(-\frac {h \nu_{10}} {k T}\right)}{1 - \exp\left(-\frac {h \nu_{10}} {k T_{0}}\right)}\right].
\end{equation}
$S'_{0}$ and $Q_{0}$ refer to the reference temperature $T_{0}$, and $S'$ and $Q$ are at the new temperature $T$. The total internal partition functions for temperatures between 70 -- 3005 K can be calculated by using the empirical equation
\begin{equation}
Q(T) = a + bT + cT^{2} + dT^{3}
\end{equation}
as given by \citet{gamache00}. The constants in this equation change depending on the temperature range and the calculated internal partition function for NH$_{3}$ is given in Table \ref{tab4}.

\begin{deluxetable}{cc}
\tabletypesize{\small}
\tablenum{4}
\label{tab4}
\tablewidth{0pt}
\tablecaption{The partition function of NH$_{3}$.}
\tablehead{\colhead{T (K)} & \colhead{Partition Function\tablenotemark{a}}}
\startdata
 296   &    1,729.30\\
 573   &    5,305.19\\
 673   &    7,218.42\\
 773   &    9,608.29\\
 873   &   12,665.85\\
 973   &   16,582.17\\
1073   &   21,548.31\\
1173   &   27,755.32\\
1273   &   35,394.26\\
1373   &   44,656.18\\
1473   &   55,732.16\\
1573   &   70,299.76\\
1643   &   81,868.17\\
\enddata
\tablenotetext{a}{Calculated using Equation 4 taken from \citet{gamache00}.}
\end{deluxetable}

\clearpage

The fitted line lists were compared with the transformed HITRAN line lists at the corresponding temperature. This allowed an intensity linear calibration factor to be determined from a comparison of our intensities to HITRAN. These calibration factors are summarized in Table \ref{tab5} and result in 24 intensity calibrated line lists (one for each temperature per region). The two regions were then combined to provide 12 line lists at each temperature.

\begin{deluxetable}{ccc}
\tabletypesize{\small}
\tablenum{5}
\label{tab5}
\tablewidth{0pt}
\tablecaption{Linear calibration factors ($f$) to convert the measured line intensities ($S_{arb}$) into HITRAN units, $S^{\prime}$ in (cm/molecule).}
\tablehead{ \multirow{2}{*}{Temperature ($^{\circ}$C)} & \colhead{Region 1\tablenotemark{a}} & \colhead{Region 2\tablenotemark{a}} \\ & \colhead{$f_1$ ($\times10^{19}$)} & \colhead{$f_2$  ($\times10^{18}$)}}
\startdata
300    &    $1.0$ &    $15.0$  \\
400    &    $1.0$ &    $10.0$  \\
500    &    $1.0$ &    $7.0$   \\
600    &    $1.0$ &    $5.0$   \\
700    &    $1.25$ &   $4.0$   \\
800    &    $1.25$ &   $3.5$   \\
900    &    $1.25$ &   $3.0$   \\
1000   &    $1.4$ &    $2.3$   \\
1100   &    $1.5$ &    $2.8$   \\
1200   &    $3.5$ &    $5.5$   \\
1300   &    $4.0$ &    $7.0$   \\
1370   &    $4.5$ &    $7.0$   \\
\enddata
\tablenotetext{a}{Where $S^{\prime}=f_{1(2)} S_{arb}$.}
\end{deluxetable}

\begin{figure}[H]
\figurenum{5}
\label{fig5}
\epsscale{1.0}
\plotone{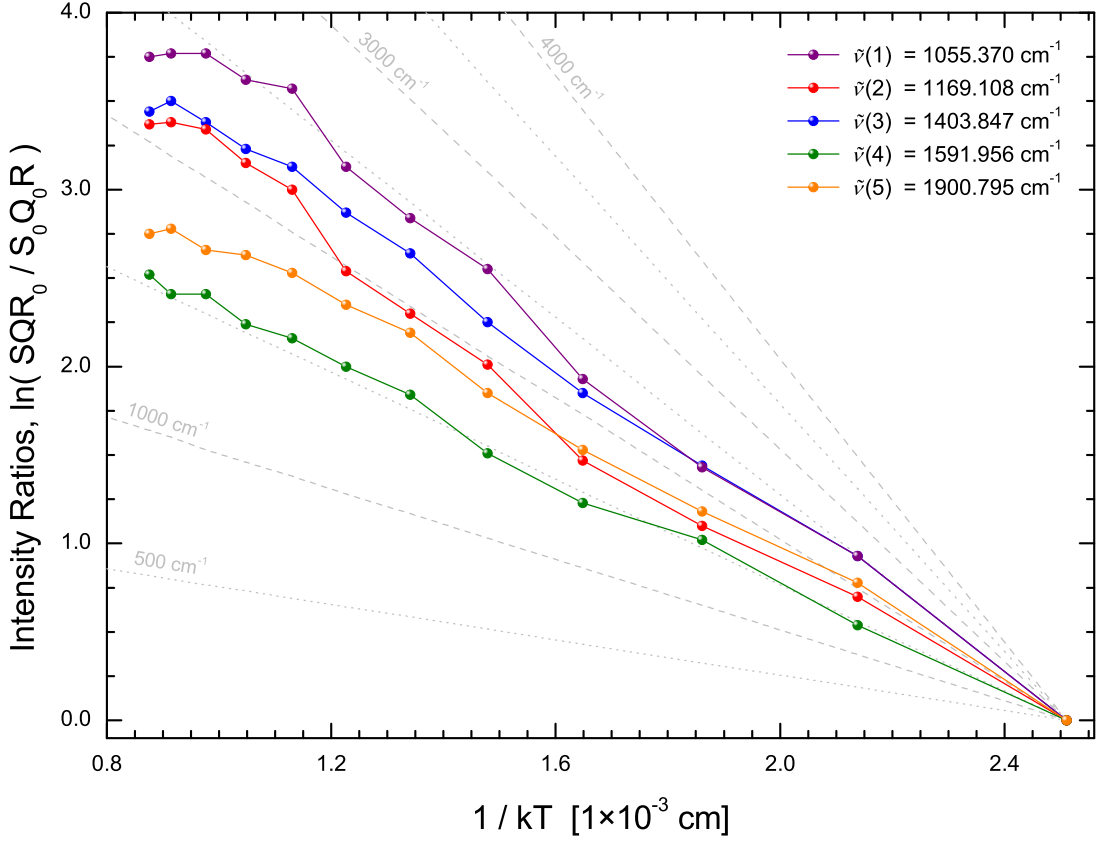}
\figcaption{The natural log of the intensity ratios of five lines present in all 12 line lists are plotted against $1/kT$ (in cm$^{-1}$), where $T$ is the temeprature. The slope of each line directly yields the empirical lower state energy and for the lines plotted, $E_{Low}(1)=2519.7$ cm$^{-1}$ [$\tilde{\nu}(1)$], $E_{Low}(2)=2091.7$ cm$^{-1}$ [$\tilde{\nu}(2)$], $E_{Low}(3)=2221.7$ cm$^{-1}$ [$\tilde{\nu}(3)$], $E_{Low}(4)=1564.0$ cm$^{-1}$ [$\tilde{\nu}(4)$], and $E_{Low}(5)=1808.6$ cm$^{-1}$ [$\tilde{\nu}(5)$]. Points above 1000$^{\circ}$C ($1/kT<0.00113$ cm$^{-1}$) were neglected in the fit.}
\end{figure}

All calibrated line lists were then compared (to within $\pm0.005$ cm$^{-1}$) to identify the same lines at every temperature. Rearranging Equation 3 and taking the natural log of both sides gives

\begin{equation}
\ln\left(\frac{SQ}{S_{0}Q_{0}} \left[\frac{1 - \exp\left(-\frac {h \nu_{10}} {k T_{0}}\right)}{1 - \exp\left(-\frac {h \nu_{10}} {k T}\right)}\right] \right) = \ln\left(\frac{SQR_{0}}{S_{0}Q_{0}R}\right) = \frac {E_{Low}} {k T_{0}} - \frac {E_{Low}} {k T},
\end{equation}

so a plot of $\ln(SQR_{0}/S_{0}Q_{0}R) $ against $1/kT$ yields a straight line. The slope of the fitted line provides the empirical lower state energy ($E_{Low}$) as illustrated with five typical lines which are present in all 12 line lists (Figure \ref{fig5}). Since all of the lines are not present in HITRAN, only the $E_{Low}$ values of $\tilde{\nu}(4)$ [$E_{Low}(4)=1564.0$ cm$^{-1}$] and $\tilde{\nu}(5)$ [$E_{Low}(5)=1808.6$ cm$^{-1}$] can be compared to HITRAN which gives an accuracy of 7\% and 2\% respectively (compared to HITRAN $E_{Low}$ values of 1673.5 and 1771.3 cm$^{-1}$). Note that we have chosen to work with the same units as in the HITRAN database. Conversions to other units, particularly for intensities (e.g. oscillator strengths, $f$ and Einstein $A$ values) can be obtained by using equations given in \citet{bernath05}.

\section{RESULTS AND ANALYSIS}

\begin{figure}[h]
\figurenum{6}
\label{fig6}
\epsscale{1.0}
\plotone{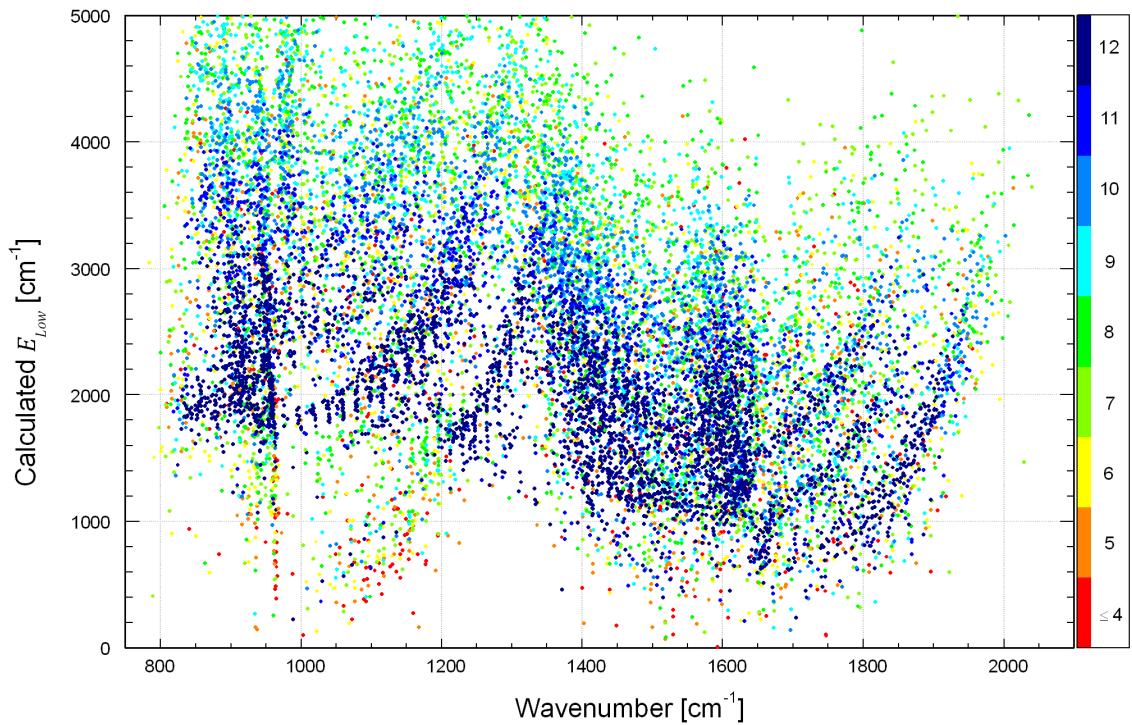}
\figcaption{Empirical lower state energies ($E_{Low}$) determined from all line lists. The colour refers to the number of temperatures at which the line appeared (12 being the maximum). It is indicative of the quality of the calculated $E_{Low}$.}
\end{figure}

\begin{figure}[h]
\figurenum{7}
\label{fig7}
\epsscale{1.0}
\plotone{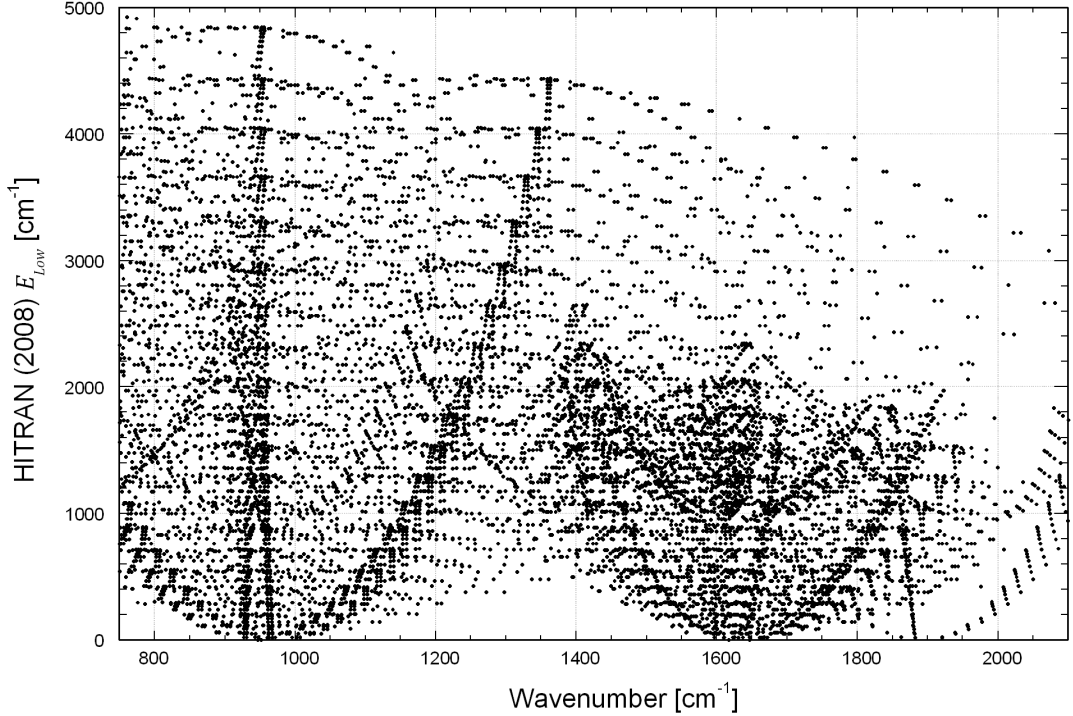}
\figcaption{Lower state energies taken from HITRAN. P- Q- and R-branches can clearly be identified in the plot.}
\end{figure}

We were able to obtain the lower state energies for the majority of the observed lines and a plot of the results in the region 740 -- 2100 cm$^{-1}$ can be seen in Figure \ref{fig6}. There is good agreement with the HITRAN lower state energies for the same region displayed in Figure \ref{fig7}, and the pattern of lines is similar. Note that because of self-absorption cold NH$_{3}$ in the end of the Al$_{2}$O$_{3}$ tube means that the strong room temperature lines are missing. They have been added back into the final line lists from HITRAN to make each line list complete.


The majority of the $\nu_{2}$ bending mode (umbrella mode) can be seen in Figures \ref{fig6} and \ref{fig7}. The band gap of the MCT detector means that we cannot detect any lines below 740 cm$^{-1}$. What is most striking about this region is the strong Q-branches rising sharply between 900 and 1000 cm$^{-1}$. Our calculated empirical lower state energies only identify these Q-branches above $E_{Low}\approx1500$ cm$^{-1}$. The patterns in Figures \ref{fig6} and \ref{fig7} can be explained qualitatively using the rigid rotor energy level formula,

\begin{equation}
E_{v,J,K}=E_{v}+BJ(J+1)+(C-B)K^{2}
\end{equation}
\begin{equation}
E_{v,J,K}[\textrm{cm}^{-1}]=E_{v}+9.94J(J+1)-3.75K^{2}
\end{equation}

with $E_{v}$ the vibrational energy. The parabolic features increasing to the right (left) of the Q-branches are due to the R-branches (P-branches) of different vibrational levels. The small nearly vertical strips within the $\nu_{2}$ region around 950 cm$^{-1}$ are due to K-structure ($\Delta J=\pm1$, $\Delta K=0$) and the doubling of the band due to inversion of the NH$_{3}$ molecule is clearly evident. Also apparent at high $E_{Low}$ in Figure \ref{fig7} are decreasing parabolic features due to forbidden K-structure ($\Delta K=\pm3$, $\Delta K=\pm6$) transitions. The R-branch in the $\nu_{2}$ region at $E_{Low}\approx1800$ cm$^{-1}$ above the ground state is likely due to the population in the $v_2=2$ vibrational level (i.e., $3\nu_{2}-2\nu_{2}$ band).

The Q-branch of the $\nu_{4}$ bending mode can also be seen rising sharply at 1650 cm$^{-1}$; the P- and R-branches are also visible as gradually increasing parabolas from the same point. The majority of lines in the HITRAN database are fundamental transitions but the $\nu_{2}+\nu_{4}-\nu_{2}$ hot band is also present at 1600 cm$^{-1}$ above an $E_{Low}\approx1000$ cm$^{-1}$. Our high temperature observations include several additional hot band transitions and a hot band is clearly visible in the $\nu_{4}$ region (Figure \ref{fig6}) with an  $E_{Low}\approx1500$ cm$^{-1}$ which can likely be assigned to the $2\nu_{4}-\nu_{4}$ hot band.

What is clearly illustrated in Figures \ref{fig6} and \ref{fig7}, is the lack of observed transitions with lower state energies below approximately $\approx1500$ cm$^{-1}$ for the $\nu_{2}$ band and $\approx1000$ cm$^{-1}$ for the $\nu_{4}$ band. The main reason for this is due to the experimental setup (shown in Figure \ref{fig2}). The NH$_{3}$ is heated at the center of the tube and emission occurs in this hot region; this emitted radiation travels along the tube to the window (maintained at room temperature). NH$_{3}$ close to the window is at approximately room temperature and absorbs some of the emitted radiation, mainly for the low \textit{J} fundamental lines. When the emission is observed, the majority of emission lines are from hot bands as well as high \textit{J} fundamental lines that are not populated at room temperature.

\begin{figure}[h]
\figurenum{8}
\label{fig8}
\epsscale{1.0}
\plotone{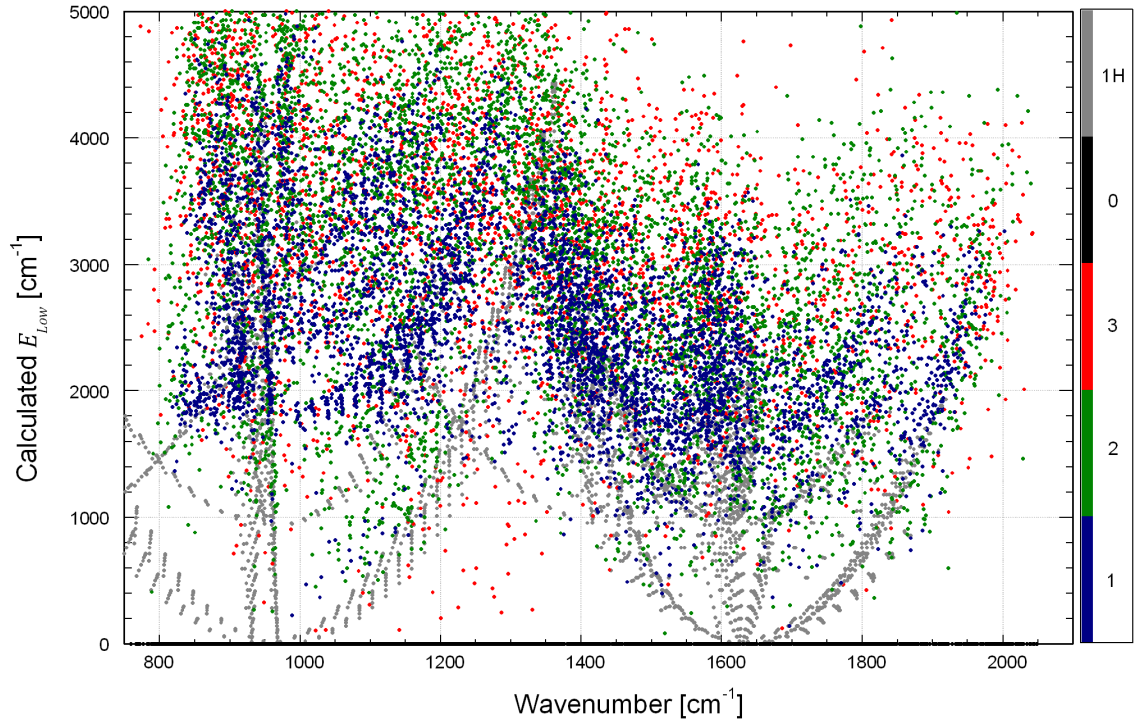}
\figcaption{Lower state energies plotted from the 1000$^{\circ}$C line list. The colour refers to the quality of the calculated empirical lower state energy.}
\end{figure}

To make our line lists complete we have inserted all HITRAN lines with an intensity above $1\times10^{-22}$ cm/molecule at each temperature. As some of the observed emission lines are due to these HITRAN lines, the observed lines have been removed to avoid the same line appearing twice. Table \ref{tab6} lists the number of lines in each file broken down into added HITRAN lines and observed lines (the total number of observed lines starts to decrease above 1100 $^{\circ}$C because of NH$_{3}$ decomposition in the cell). The consequence of this procedure can be seen from a lower state energy plot of the 1000$^{\circ}$C line list (Figure \ref{fig8}). The 12 line lists at each temperature have been amalgamated into one table which is available in a machine readable version online. A sample of this final line list for all temperatures is provided in Table \ref{tab7}.

\begin{deluxetable}{cccc}
\tabletypesize{\small}
\tablenum{6}
\label{tab6}
\tablewidth{0pt}
\tablecaption{The number of lines present in each line list.}
\tablehead{ \colhead{Temperature ($^{\circ}$C)} & \colhead{Total Lines} & \colhead{Observed Lines} & \colhead{Added HITRAN Lines}}
\startdata
300    &    8,102 &    3,611   &    4,491 \\
400    &   10,264 &    3,758   &    6,506 \\
500    &   14,043 &    3,816   &   10,227 \\
600    &   16,305 &    3,784   &   12,521 \\
700    &   19,965 &    3,688   &   16,277 \\
800    &   19,003 &    3,574   &   15,429 \\
900    &   22,680 &    3,431   &   19,249 \\
1000   &   24,168 &    3,295   &   20,873 \\
1100   &   24,506 &    3,173   &   21,333 \\
1200   &   23,151 &    3,084   &   20,067 \\
1300   &   20,089 &    2,955   &   17,134 \\
1370   &   15,410 &    2,873   &   12,537 \\
\enddata
\end{deluxetable}

\clearpage

\begin{deluxetable}{ccccc}
\tabletypesize{\small}
\rotate{90}
\tablenum{7}
\label{tab7}
\tablewidth{0pt}
\tablecaption{A sample of 10 lines from the 1000$^{\circ}$C line list.}
\tablehead{ \colhead{Temperature ($^{\circ}$C)} & \colhead{Wavenumber (cm$^{-1}$)} & \colhead{Intensity (cm/molecule)} & \colhead{$E_{Low}$ (cm$^{-1}$)} & \colhead{Quality Identifier} }
\startdata
... & ... & ... & ... & ... \\
1000& 1591.8243452&  0.68853731E-22&  0.15846541E+04&  2 \\
1000& 1591.8273898&  0.68853731E-22&  0.00000000E+00&  0 \\
1000& 1591.8536200&  0.62000000E-21&  0.14309118E+04&  1H \\
1000& 1591.9071469&  0.36946435E-21&  0.26669918E+04&  2 \\
1000& 1591.9233494&  0.42613992E-21&  0.84666645E+03&  3 \\
1000& 1591.9558781&  0.41360202E-21&  0.15639455E+04&  1 \\
1000& 1591.9876537&  0.16819148E-21&  0.33495921E+04&  2 \\
1000& 1592.0143421&  0.75043627E-22&  0.30381832E+04&  3 \\
1000& 1592.0369615&  0.79134180E-22&  0.15844298E+04&  2 \\
1000& 1592.0767117&  0.97189882E-21&  0.16858822E+04&  1 \\
... & ... & ... & ... & ... \\
\enddata
\end{deluxetable}
\clearpage

The estimation of the error in the empirical lower state energies is difficult. As a measure of the accuracy of our empirically-determined lower state energies, we determined the differences between our values and those given in HITRAN for the lines that are present in both data sets. The average $E_{low}$ difference was found to be 147 cm$^{-1}$ which corresponds to 3.5\%. However, the average magnitude of the deviation is determined to be 364 cm$^{-1}$ corresponding to 30\%. Since these lines have been largely removed from the final line lists and replaced by the actual HITRAN values due to self absorption issues, we consider these errors to be an over estimation.

Another measure of the quality of the empirical lower state energy can be obtained from the number of points (i.e. the number of line lists each particular line appears in) used to calculate the value. For lines occurring in 10 -- 12 spectra the $E_{Low}$ was calculated from the points below 1000$^{\circ}$C and has a quality factor of `1', these are considered to have the least error. For lines in which an $E_{Low}$ can still be determined from points below 1000$^{\circ}$C (i.e. occur in at least 3 spectra below 1000$^{\circ}$C) but occur in less than 10 spectra overall, the quality factor is `2'. For any line in which an $E_{Low}$ is still able to be determined but only from 3 points including spectra above 1000$^{\circ}$C, then the quality factor is `3', and are considered to have the greatest error. Any line where $E_{Low}$ cannot be accurately determined (i.e. less than 3 points), then the quality factor is `0'. The HITRAN lines which occur in the line lists have a quality factor of `1H' and are are taken from \citet{rothman09}. The quality factors can be seen in Table \ref{tab7} and the effect on the lower state energy can be seen in Figure \ref{fig8}.

\section{DISCUSSION}

The ammonia line lists have been wavenumber calibrated with the HITRAN 2008 database and are accurate to within $\pm$0.002 cm$^{-1}$. The intensity was also calibrated with the HITRAN 2008 database by extrapolating the intensities using Equation 3 to the temperatures of the recorded spectra. An error in the calibrated intensity of perhaps a factor of two is suggested for our line lists. Although this seems large, emission intensities are notoriously difficult to calibrate \citep{nassar03}.

\begin{figure}[H]
\figurenum{9}
\label{fig9}
\epsscale{1.0}
\plotone{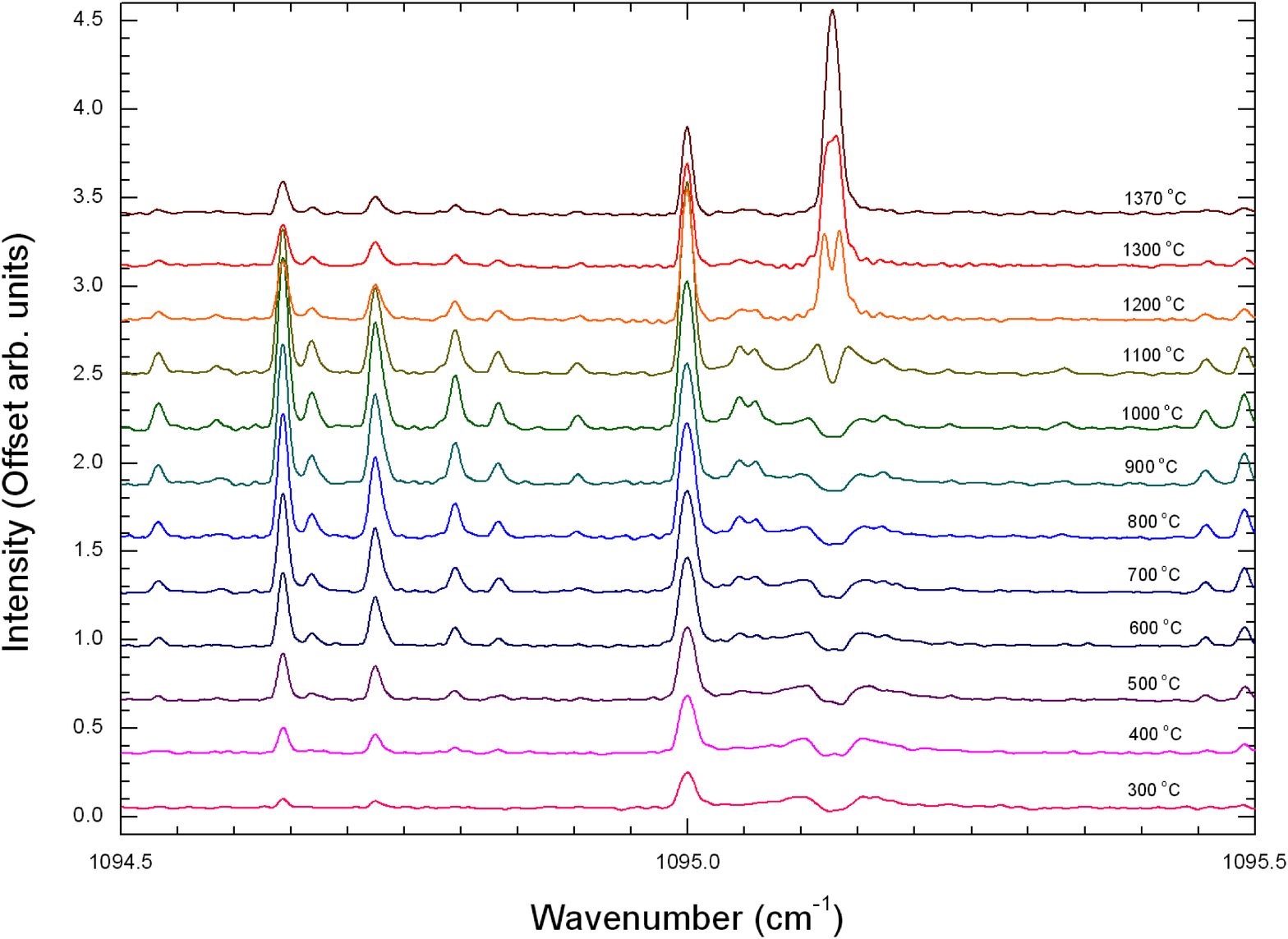}
\figcaption{A short (1 cm$^{-1}$) section of the NH$_{3}$ emission spectra. Notice that the line at 1095.13 cm$^{-1}$ (R(7) of the fundamental $\nu_{2}$ band) changes from absorption to emission as the temperature increases.}
\end{figure}

We have accounted for the instrument response by recording a blackbody spectrum and correcting the intensities with the instrument response function. The most notable corrections were obvious towards the edge of each spectral region (see Figure \ref{fig4}). Blending of lines also causes errors in line intensities and in the empirical lower state energy. These blended lines cause problems when matching our observations with HITRAN and make it difficult to identify the contribution of each line to the observed peak. We have tried to reduce our intensity calibration error by comparing as many spectral lines with HITRAN lines as possible. Each comparison gives an intensity calibration factor and after outliers have been removed a general factor can be obtained. The temperature of the experiment was accurate to within $\pm$10 K but this only applies to the central portion of the Al$_{2}$O$_{3}$ tube which is surrounded by the furnace. There is therefore a temperature gradient within the Al$_{2}$O$_{3}$ tube which affects the shape of some lines as discussed by \citet{nassar03}. This effect can be observed in the emission spectra as we see absorption of the low \textit{J} fundamental NH$_{3}$ lines due to cooler gas towards the tube ends change to emission as the temperature of the tube increases (see Figure \ref{fig9}).

At higher temperatures the Al$_{2}$O$_{3}$ tube ends are sufficiently hot and the emission from the center is stronger so they no longer completely absorb the low \textit{J} fundamental lines leading to differences between successive spectra starting around 1000 $^{\circ}$C. As long as the center of the tube is `cool' ($<$900 $^{\circ}$C) the fundamental low \textit{J} lines appear in absorption, however for temperatures greater than 1000 $^{\circ}$C the fundamental low \textit{J} lines start to appear in emission. Moreover, because the Doppler broadening depends on temperature the emitted lines are broader than the absorption lines. This leads to an absorption feature in the middle of an emission line which can sometimes cause apparent splitting of the line (e.g. the line at 1095.13 cm$^{-1}$ in Figure \ref{fig9}). To solve this problem, we have processed each line list and inserted HITRAN lines with intensities greater than $1\times10^{-22}$ cm/molecule at that temperature, denoted by `1H' in the line lists. We also simultaneously removed the emission lines showing this splitting effect, and also any lines which are within $\pm$0.005 cm$^{-1}$ of each HITRAN line with an intensity greater than $1\times10^{-22}$ cm/molecule. Our line lists therefore provide a complete representation of NH$_{3}$ between 740 -- 2100 cm$^{-1}$ at temperatures between 300 -- 1370$^{\circ}$C.

A useful check for the quality of the calibrated intensities determined between 740 -- 2100 cm$^{-1}$ can be verified by an intensity sum comparison with the temperature scaled HITRAN intensities \citep{nassar03}. At 300$^{\circ}$C, the sum of the intensities in our linelist is only 7\% greater than the sum of the intensity scaled HITRAN lines but as the temperature increases we continue to exceed the HITRAN total intensity with a maximum of 105\% greater at 1200$^{\circ}$C (Table \ref{tab8}). We attribute our larger values to an increase in the number of hot band lines present in the spectrum. These hot lines only have a small contribution at `low' temperatures (300$^{\circ}$C) but are more significant for higher temperatures ($>1000^{\circ}$C) as hot transitions become more favorable. Indeed, this comes as no surprise because the HITRAN database consists mainly of fundamental transitions and highlights how our line lists are more suitable for astrophysical applications.

\begin{deluxetable}{cccc}
\tabletypesize{\small}
\tablenum{8}
\label{tab8}
\tablewidth{0pt}
\tablecaption{Intensity sums for each line list and HITRAN between 740--2100 cm$^{-1}$.}
\tablehead{ \colhead{Temperature} & \colhead{Intensity Sum\tablenotemark{a}} & \colhead{HITRAN intensity sum} & \colhead{Ratio to} \\ \colhead{($^{\circ}$C)} & \colhead{(cm$^{-1}$)} & \colhead{(cm$^{-1}$)} & \colhead{HITRAN} }
\startdata
300    &   $2.70\times10^{-17}$ &   -   &    - \\
300    &   $2.24\times10^{-17}$ &    $2.39\times10^{-17}$   &   1.07 \\
400    &   $2.04\times10^{-17}$ &    $2.27\times10^{-17}$   &   1.11 \\
500    &   $1.82\times10^{-17}$ &    $2.12\times10^{-17}$   &   1.16 \\
600    &   $1.59\times10^{-17}$ &    $1.95\times10^{-17}$   &   1.23 \\
700    &   $1.37\times10^{-17}$ &    $1.85\times10^{-17}$   &   1.35 \\
800    &   $1.17\times10^{-17}$ &    $1.68\times10^{-17}$   &   1.44 \\
900    &   $9.89\times10^{-18}$ &    $1.54\times10^{-17}$   &   1.56 \\
1000   &   $8.34\times10^{-18}$ &    $1.49\times10^{-17}$   &   1.77 \\
1100   &   $7.03\times10^{-18}$ &    $1.33\times10^{-17}$   &   1.89 \\
1200   &   $5.94\times10^{-18}$ &    $1.22\times10^{-17}$   &   2.05 \\
1300   &   $4.92\times10^{-18}$ &    $9.64\times10^{-18}$   &   1.96 \\
1370   &   $4.35\times10^{-18}$ &    $8.05\times10^{-18}$   &   1.85 \\
\enddata
\tablenotetext{a}{The intensity sums were calculated before the removal and addition of lines from the calibrated line lists.}
\end{deluxetable}

\clearpage

A similar method to determine empirical lower state energies has been used by \citet{fortman10} by recording numerous millimeter wave spectra with the Fast Scan Submillimeter Spectroscopic Technique (FASST) as a function of temperature. For NH$_{3}$, empirical lower state energies were also determined by \citet{brownmarg96} for unassigned lines in the 4791 -- 5294 cm$^{-1}$ region. Another method has been used recently to obtain lower state energies for near infrared lines of CH$_{4}$ by cooling an absorption cell \citep{wang10}. For the FT-IR method presented here, we have line lists at 12 individual temperatures each of which requires many hours to obtain. As a result, we are only able to use a maximum of 12 points in the calculation of the empirical lower state energy for each line and in fact, this was reduced to 8 points as the intensity of the lines above 1000$^{\circ}$C begin to decrease in strength due to the dissociation of the molecule. Our accuracy depends on the number of points used in the calculation of $E_{Low}$ and for this reason, our line lists also include a quality factor which only applies to the empirical lower state energy. This $E_{Low}$ method is used to determine an important characteristic of each line whilst knowing nothing of the spectroscopic quantum number assignments.

The inclusion of HITRAN lines in our line lists has benefits as well as problems. The main benefit is that the line lists now contain the strong lines which were absent from the observations and as a result are more suitable for simulation of astronomical spectra (Figure \ref{fig1}). As a consequence of including these strong lines ($1\times10^{-22}$ cm/molecule), some observed lines have been removed because they have the same frequency as the HITRAN line. It was decided that the benefits of including the HITRAN lines outweighed the costs of removing coincident emission lines.

The errors associated with the empirical lower state energies have been based on comparisons with HITRAN however, as the majority of lines used to determine the errors are self absorbed, it does not accurately represent all of the lines in each spectrum. Indeed, at low temperatures the strong lines which are present display self absorption even if it is not immediately apparent from the line shape and at high temperatures there are few lines to draw conclusive comparisons from, since they are not given in HITRAN. The overall maximum error of the empirical lower state energies in the line list should therefore be taken as 30\%, but we believe that the quality factor gives a better indication of how accurate the empirical lower state energy is for each line.

There are two approaches to obtaining an ammonia line list to model brown dwarf and exoplanet spectra: the first is experimental measurement consistent with the work presented here and the second is an \textit{ab initio} theoretical calculation. These two methods each have their own strengths and weaknesses, but in general are complementary. The experimental approach provides fewer lines but more accurate line positions, while the theoretical approach provides many more lines but with reduced line position accuracy. Calculated line lists are particularly useful for assigning quantum numbers to the measured lines, which can then be used to improve the calculations. Such line assignments of new NH$_{3}$ data are currently underway and will be reported elsewhere \citep{zobov11}.

The line lists provided are given in the format shown in Table \ref{tab7} and can be obtained as electronic supplements with extra details and spectra provided upon request. It is important to note that the line lists are temperature specific and should only be used at the appropriate temperature. If the intensities need to be scaled, then the closest temperature line list should be chosen and the empirical lower state energies can be used to adjust the lines intensities as needed (using Equation 3) to reach additional temperatures.

We are currently in the process of completing our NH$_{3}$ measurements in the 3 micron region and are set to record new spectra in the near infrared; a similar comprehensive study of CH$_{4}$ is also underway.

\section{CONCLUSION}

We have produced `hot' experimental line lists of NH$_{3}$ at temperatures 300 -- 1300$^{\circ}$C in 100$^{\circ}$C intervals and at 1370$^{\circ}$C. Each line list has been wavenumber and intensity calibrated using the HITRAN 2008 database and from this we have been able to derive empirical lower state energies from the temperature dependance of line intensities. The line lists can be used directly to model NH$_{3}$ between 740 -- 2100 cm$^{-1}$ (4.76 -- 13.5 $\mu$m) in brown dwarfs and exoplanets.

\acknowledgments

Support for this work was provided by a Research Project Grant from the Leverhulme Trust and a Department of Chemistry (University of York) studentship.

\end{document}